\documentstyle[aps,prl,floats,epsf]{revtex}

\newcommand{\bastar}{\begin{eqnarray*}}
\newcommand{\eastar}{\end{eqnarray*}}
\newskip\humongous \humongous=0pt plus 1000pt minus 1000pt

\newif\ifdtup

\relax
\newcommand{\be}{\begin{equation}}
\newcommand{\ee}{\end{equation}}
\newcommand{\bea}{\begin{eqnarray}}
\newcommand{\eea}{\end{eqnarray}}

\newcommand{\tF}{{\tilde F}}

\newcommand{\pro}{\partial}

\newcommand{\dfrac}{\displaystyle\frac}
\newcommand{\ba}{\begin{array}}
\newcommand{\ea}{\end{array}}

\newcommand{\nn}{\nonumber}

\newcommand{\Ei}{{\rm Ei}}
\newcommand{\si}{{\rm si}}
\newcommand{\ci}{{\rm ci}}
\newcommand{\Eibar}{{\overline {\rm Ei}}}
\begin{document}
\twocolumn[\hsize\textwidth\columnwidth\hsize\csname@twocolumnfalse%
\endcsname
\title  {Electric-Magnetic Duality in QED Effective Action}
\bigskip
\author{W. S. Bae and Y. M. Cho}
\address{
Department of Physics, College of Natural Sciences, Seoul National University,
Seoul 151-742, Korea\\
{\scriptsize  hypersym@zoo.snu.ac.kr, ymcho@yongmin.snu.ac.kr}\\
~~\\
{\normalsize\rm and}}
\author {D. G. Pak}
\address{
Asia Pacific Center for Theoretical Physics, 207-43 Cheongryangri-dong, Dongdaemun-gu,
               Seoul 130-012 Korea \\
{\scriptsize dmipak@mail.apctp.org} \\ \vskip 0.3cm }
\maketitle

\begin{abstract}

Recently we have obtained a non-perturbative
but convergent series expression of the one loop effective action of QED, and
discussed the renormalization of the effective action.
In this paper we establish the electric-magnetic duality
in the quantum effective action.

\vspace{0.3cm}
\end{abstract}

\narrowtext
\bigskip
                           ]

The effective theory of QED plays a crucial role in our understanding of the non-linear effects in
electrodynamics. Recently we have obtained a convergent series expression of the effective action of QED in one
loop approximation,and established the renormalization group invariance of the effective action  \cite{cho1}. A
remarkable feature of the effective action is the electric-magnetic duality,a fundamental symmetry of the quantum
effective action of QED. The purpose of this paper is to provide a concrete proof of the duality in the effective
action of QED.\par
The effective action of QED has been studied by Euler and Heisenberg and by Schwinger long
time ago \cite{eul,sch},and by many others later \cite{ditt,ritus}.To derive the effective action one may start
from the QED Lagrangian
\bea
& {\cal L} = -\dfrac{1}{4} F_{\mu \nu}^2 + \bar \Psi( i {{/ \,}\llap D} -m) \Psi, \nn\\& D_\mu= \pro_\mu
+ieA_\mu,
\eea
where $m$ is the electron mass.With a proper gauge fixing one can show that the electron one loop correction
of the effective action is given by
\bea
\Delta S &=& i \ln {\rm Det} (i {{/ \,} \llap D}  - m) .
\eea
So for an arbitrary constant background one has \cite{eul,sch}
\bea
\Delta {\cal L} = - \dfrac{ab}{8\pi^2}\int_{0+i\epsilon}^{\infty+i\epsilon} \dfrac{dt}{t} \coth (at) \cot (bt)
\exp(-m^2 t),
\eea
where
\bea
a = \dfrac{e}{2} \sqrt {\sqrt {F^4 + (F \tF)^2} + F^2}, \nn\\
b = \dfrac{e}{2} \sqrt {\sqrt {F^4 + (F \tF)^2} - F^2}. \nn
\eea
Notice that the above contour of the integral is dictated by the causality. Although the integral expression (3)
looks innocent, it has been non-trivial to perform the integral for an arbitrary constant background
\cite{ditt,ritus}.But we can integrate this and obtain a convergent series expression of the effective action with
the help of the Sitaramachandrarao's identity \cite{sita},
\bea
& xy \coth x \cot y = 1 + \dfrac{1}{3} (x^2-y^2)  \nn \\
& -  \dfrac{2}{\pi}  x^3 y  \sum_{n=1}^{\infty} \dfrac{1}{n} \dfrac{\coth(\dfrac{n \pi y}{x})}{(x^2 +n^2  \pi^2)}
\nn \\
&  +\dfrac{2}{\pi}    x y^3  \sum_{n=1}^{\infty} \dfrac{1}{n} \dfrac{\coth (\dfrac{n \pi x}{y})}{(y^2 - n^2
\pi^2)}.
\eea
A preliminary version of the identity which was expressed as a divergent asymptotic series was obtained by
Ramanujan, and Sitaramachandrarao improved the identity in the above convergent series expression. To carry out
the integral (3)and express it as a convergent series expression we need the Sitaramachandrarao's identity. Indeed
with the identity we obtain the following one loop effective action for an arbitrary constant background (after the
modified minimal subtraction)\cite{cho1,miel}
\bea
&&{\cal L}_{eff} =-\dfrac{a^2-b^2}{2e^2}{\Big (}1-\dfrac{e^2}{12\pi^2}\ln\dfrac{m^2}{\mu^2}{\Big)} \nn\\
&& -\dfrac{ab}{4\pi^3}\sum_{n=1}^{\infty}\dfrac{1}{n}{\Big [}\coth(\dfrac{n \pi b}{a}){\Big (} {\rm ci}(\dfrac{n
\pi m^2}{a})\cos(\dfrac{n \pi m^2}{a}) \nn\\
&& +{\rm si}(\dfrac{n \pi m^2}{a}) \sin(\dfrac{n \pi m^2}{a}){\Big )} \nn\\
&&-\dfrac{1}{2} \coth (\dfrac{n \pi a}{b}) {\Big (} \exp(\dfrac{n \pi m^2}{b}){\rm Ei}(-\dfrac{n \pi m^2}{b})
\nn \\
&&+ \exp(-\dfrac{n \pi m^2}{b}){\rm Ei}(\dfrac{n \pi m^2}{b}-i \epsilon){\Big )}{\Big ]},
\eea
where $\mu$ is the subtraction parameter, and ci($x$), si($x$), and  Ei($-x$) are the cosine, sine, and
exponential integral functions given by (for Re $x > 0$) \cite{abra},
\bea
{\rm ci}(x)&=&-\int_{x}^{\infty} \dfrac{\cos(t)}{t}  dt \nn\\
&=& \gamma + \ln x + \sum_{n=1}^{\infty}\dfrac{(-1)^n x^{2n}}{(2n)!2n}, \nn \\
{\rm si}(x)&=&-\int_{x}^{\infty} \dfrac{\sin(t)}{t} dt \nn\\
&=&-\dfrac{\pi}{2}-\sum_{n=1}^{\infty}\dfrac{(-1)^n x^{2n-1}}{(2n-1)!(2n-1)}, \nn\\
{\rm Ei}(-x)&=&-\int_{x}^{\infty} \dfrac{e^{-t}}{t} dt \nn\\
&=&\gamma+\ln x +\sum_{n=1}^{\infty} \dfrac{(-1)^n x^{n}}{(n)!n}.
\eea
Notice that our series expression of the effective action is non-perturbative but
convergent.\par
The effective action
(5) nicely reproduces the well-known imaginary part when $b \neq 0$ \cite{sch,ditt},
\bea
{\rm Im}~{\cal L}_{eff} = \dfrac{ab}{8 \pi^2 } \sum_{n=1}^{\infty}    \dfrac{1}{n}   \coth (\dfrac{n \pi a}{b})\,
\exp(-\dfrac{n \pi m^2}{b}).
\eea
This is because the exponential integral function ${\rm Ei}(-x)$ in (5) develops an imaginary part $i\pi$ after
the analytic continuation from $-x$ to $x$ \cite{cho1}. The important point here is that one should make the
analytic continuation in such a way to preserve the causality, which determines the signature of the imaginary
part in (7). The physical meaning of the imaginary part is well known \cite{sch}. The electric background
generates the pair creation,with the probability per unit volume per unit time given by (7).

Another important aspect of the effective action is the logarithmic correction $\ln (m/\mu)^2$
of the classical part of the action. This indicates that even the classical part of the action
gets the quantum correction. One might think that this correction is unphysical and should disappear
after the renormalization, since one can remove this logarithmic
correction by choosing $\mu=m$. But this is not true. Indeed it has been shown that there exists a real physical
quantum correction to the classical part of the action even after the renormalization
\cite{cho1}.\par
Notice that in the pure magnetic and the pure electric background the effective action (5) reduces to
\bea
&{\cal L}_{eff} = - \dfrac{a^2 }{2e^2}{\Big (}1-\dfrac{e^2}{12\pi^2} \ln\dfrac{m^2}{\mu^2}{\Big)} \nn\\
& - \dfrac{a^2}{4\pi^4}\sum_{n=1}^{\infty}\dfrac{1}{n^2}{\Big (} {\rm ci}(\dfrac{n \pi m^2}{a}) \cos(\dfrac{n \pi
m^2}{a}) \nn\\
& + {\rm si}(\dfrac{n \pi m^2}{a}) \sin (\dfrac{n \pi m^2}{a}) {\Big )},
\eea
and to
\bea
&{\cal L}_{eff} = \dfrac{ b^2}{2e^2} {\Big (}1-\dfrac{e^2}{12\pi^2} \ln\dfrac{m^2}{\mu^2}{\Big )} \nn\\
&+ \dfrac{b^2}{8 \pi^4}\sum_{n=1}^{\infty}\dfrac{1}{n^2}{\Big (}  \exp(\dfrac{n \pi m^2}{b}) {\rm Ei}(-\dfrac{n
\pi m^2}{b}) \nn\\
&+ \exp(-\dfrac{n \pi m^2}{b} ) {\Eibar}(\dfrac{n \pi m^2}{b} ){\Big)}\nn\\
&+ i\dfrac{b^2}{8 \pi^3}\sum_{n=1}^{\infty}\dfrac{1}{n^2}\exp(-\dfrac{n \pi m^2}{b}),
\eea
where
\bea
\overline {\Ei} (x) &=& \dfrac{1}{2}{\Big (} \Ei (x + i \epsilon) +\Ei (x - i \epsilon) \Big ) \nn \\
                &=& {\rm Re} ~\Ei (x). ~~~~~~~~~~(x>0)
\eea
Although the above effective actions for the pure electric and the pure magnetic backgrounds look totally
different, they are actually very closely related. In fact one can show that they are the mirror image of each
other, so that one can obtain one from the other simply by making a dual transformation \cite{cho1}. This is a
first indication that the quantum effective action has a fundamental symmetry which we call the duality.{\it In
general the duality asserts that the effective action of QED is manifestly invariant under the dual
transformation,}
\bea
a \rightarrow -ib,~~~~b \rightarrow  ia.
\eea
{\it This means that the effective action, as a function of $z=a+ib$, is invariant under the reflection from $z$ to
$-z$}.  Notice that, in the Lorentz frame where $\vec E$ is parallel to $\vec B$, $a$ becomes$B$ and $b$
becomes $E$. So the duality describes the electric-magnetic
duality.\par
One might think that the duality is obvious
since it immediately follows from the integral expression (3). But we emphasize that this is not so. In fact the
integral expression is invariant under the four different transformations,
\bea
a \rightarrow \pm ~ib,~~~~b \rightarrow  \pm ~/\mp ia.
\eea
But among the four only our duality (11) survives as the true symmetry of the effective action. This tells that
the duality constitutes a non-trivial symmetry of the quantum effective
action.\par
To establish the duality it is
important to realize that the dual transformation automatically involves the analytic continuation of the special
functions ci($x$), si($x$), and Ei($-x$) in (5).So one must find the correct analytic continuation to establish the
duality in the effective action.To do this we observe from (6) that \cite{abra}
\bea
\ci (x) &=& \dfrac{1}{2}(\Ei (i x) + \Ei (-ix))  \nn \\
\si (x) &=& \dfrac{1}{2i}(\Ei (i x) - \Ei (-ix)).
\eea
Notice that although ci($x$) and si($x$) look like an even and odd function respectively, they are not.
This is because Ei($x$) has the branch cut along the positive real axis. From (13) we have for a positive real $x$
\bea
&\ci (\pm ix) = \dfrac{1}{2}(\Ei (-x) + \Eibar (x))\pm i \dfrac{\pi}{2}  \nn \\
&\si (\pm ix) = \pm \dfrac{1}{2i}(\Ei (- x) - \Eibar (x))-\dfrac{\pi}{2},
\eea
and
\bea
\Ei (\pm ix) = \ci (x) \pm i ~\si (x).
\eea
Furthermore from (10) and (13) we have (for a positive real $x$)
\bea
\Eibar (ix) &=& \dfrac{1}{2} {\Big (} \Ei(ix)+ \Ei^{(+)}(ix) {\Big )} \nn\\
&=& \Ei (ix) + i \pi  \nn \\
\Eibar (-ix) &=& \dfrac{1}{2} {\Big (} \Ei^{(-)}(-ix)+ \Ei (-ix) {\Big )} \nn\\
&=& \Ei (-ix) - i \pi ,
\eea
where $\Ei^{(\pm)} (\pm ix)$ are $\Ei (\pm ix)$on the $(\pm 1)$-th Riemann
sheets.\par
With the above
preliminaries we can now establish the duality.To do this let us express the effective action (5) as
\bea
&{\cal L}_{eff} =-\dfrac{a^2-b^2}{2e^2}{\Big (}1-\dfrac{e^2}{12\pi^2}\ln\dfrac{m^2}{\mu^2}{\Big)} \nn\\
& -\dfrac{ab}{4 \pi^3}\sum_{n=1}^{\infty}\dfrac{1}{n} {\Big [}     \coth (\dfrac{n \pi b}{a} )  f_1 (nx) \nn\\
&- \coth (\dfrac{n \pi a}{b}) f_2 (ny) {\Big ]},
\eea
where
\bea
&f_1 (nx ) = \ci (nx) \cos (nx) + \si (nx ) \sin (nx), \nn \\
&f_2 (ny) = \dfrac{1}{2} {\Big (} \Ei (-ny) \exp(ny)\nn\\
&+ {\big(}\Eibar (ny)+ i \pi {\big)}\exp(-ny) {\Big)},\nn \\
&x = \dfrac{ \pi m^2}{a}, ~~~~~~~y = \dfrac{\pi m^2}{b}. \nn
\eea
Now under
\bea
a \rightarrow \mp i b  ~~~~{\text or}~~~~x \rightarrow \pm i y
\eea
we have from (14)
\bea
&f_1 (nx) \rightarrow {\Big (}\dfrac{1}{2}(\Ei (-ny)+ \Eibar (ny)) \pm i \dfrac{\pi}{2} {\Big )} \cosh (ny) \nn \\
&+ {\Big (}  \dfrac{1}{2}(\Ei (-ny) - \Eibar (ny))\mp i \dfrac{\pi}{2}{\Big )}            \sinh (ny) \nn \\
&= \dfrac{1}{2} {\Big (} \Ei (-ny) \exp(ny) \nn\\
&+{\big(}\Eibar (ny)\pm i \pi {\big)}\exp(-ny) {\Big)}.
\eea
From this we have
\bea
&f_1 (nx) \rightarrow f_2 (ny)~~for~~a \rightarrow -ib ,\nn \\
&f_1 (nx) \rightarrow f_2 (ny)-i \pi \exp(-ny)~~for~~a \rightarrow +ib.
\eea
On the other hand under
\bea
b \rightarrow \pm i a   ~~~~~{\text  or} ~~~y \rightarrow \mp ix
\eea
we have from (15) and (16)
\bea
&f_2 (ny) \rightarrow \dfrac{1}{2}{\Big (}\ci(nx)\pm i~\si(nx){\Big )}\nn\\
&{\Big (} \cos (nx) \mp i \sin (nx){\Big)} \nn \\
&+ \dfrac{1}{2} {\Big (} \ci (x) \mp i~\si (x) \mp i \pi + i \pi {\Big)}\nn\\
&{\Big (} \cos (nx) \pm i \sin (nx) {\Big )} \nn \\
&= {\Big (} \ci (nx) + \dfrac{i}{2}(\pi \mp \pi){\Big )} \cos (nx) \nn \\
&+ \Big ( \si (nx) + \dfrac {1}{2}(\pi \mp \pi) {\Big )} \sin (nx) .
\eea
From this we have
\bea
&f_2 (ny) \rightarrow f_1 (nx)~for~b \rightarrow +ia \nn \\
&f_2 (ny) \rightarrow f_1 (nx)+i \pi \exp(-inx)~for~b \rightarrow -ia.
\eea
From these it must become clear that the effective action is invariant under the dual transformation (11). As
importantly the above analysis shows that the effective action is {\it not} invariant under the other three
transformations of (12). This establishes the duality in the quantum effective action in QED.It is really
remarkable that the quantum effective action is invariant only under our duality (11), even though the integral
expression (3) appears to be invariant under the four different transformations
(12).\par
One can obtain the similar
results for the scalar QED. In this case the integral expression of the one loop correction is given by
\cite{sch,ditt}
\bea
\Delta {\cal L}_0 = \dfrac {ab}{16 \pi^2}\int_{0+i\epsilon}^{\infty+i\epsilon} \dfrac{dt}{t}{\rm csch}(at){\rm
csc}(bt)\exp(- m^2 t).
\eea
To perform the integral we need a new identity similar to the Sitaramachandrarao's identity (4)
\bea
& xy {\rm csch} x \csc y  = 1 - \dfrac{1}{6} (x^2 - y^2)  \nn \\
& - \dfrac{2}{\pi} x^3 y  \sum_{n=1}^\infty \dfrac{(-1)^n}{n} \dfrac{{\rm csch}(\dfrac{n \pi y}{x})}
{x^2 + n^2 \pi^2 } \nn \\
& + \dfrac{2}{\pi} x y^3 \sum_{n=1}^\infty \dfrac{(-1)^n}{n}\dfrac{{\rm csch} (\dfrac{n \pi x}{y})}
{y^2 - n^2 \pi^2 }.
\eea
Again a preliminary version which was divergent was obtained by Ramanujan \cite{sita}, but we have improved
the Ramanujan's identity and obtained the above convergent series \cite{cho1}. With this identity we can integrate
(24) and express the effective action as (with the modified minimal subtraction)
\bea
&{\cal L}_{0\,eff} =-\dfrac{a^2-b^2}{2e^2}{\Big (}1-\dfrac{e^2}{48\pi^2}\ln\dfrac{m^2}{\mu^2}{\Big)}\nn\\
&+\dfrac{ab}{8 \pi^3}\sum_{n=1}^{\infty} \dfrac{(-1)^n}{n}{\Big [} {\rm csch} (\dfrac{n \pi b}{a} )  f_1 (nx)\nn\\
&-{\rm csch}(\dfrac{n \pi a}{b}) f_2 (ny) {\Big ]}.
\eea
With this it is now straightforward to establish the duality in the effective action of the scalar QED. Indeed from
(20) and (23) it must be clear that the effective action is invariant under the dual transformation (11). Notice
again that the integral expression of the effective action (24) is invariant under the four different transformations
(12), but the quantum effective action (26) is invariant under only our duality
(11).\par
In the pure magnetic
background the one loop effective action (26) reduces to \cite{cho1}
\bea
&{\cal L}_{0\,eff} = - \dfrac{a^2 }{2e^2}{\Big (}1-\dfrac{e^2}{48\pi^2} \ln\dfrac{m^2}{\mu^2}{\Big)} \nn\\
& + \dfrac{a^2}{8\pi^4} \sum_{n=1}^{\infty}\dfrac{(-1)^n}{n^2}{\Big (} {\rm ci}(\dfrac{n \pi m^2}{a}) \cos(\dfrac{n
\pi m^2}{a}) \nn\\
& + {\rm si}(\dfrac{n \pi m^2}{a}) \sin (\dfrac{n \pi m^2}{a}) {\Big )},
\eea
and to
\bea
&{\cal L}_{0\,eff} =  \dfrac{ b^2}{2e^2} {\Big (}1-\dfrac{e^2}{48\pi^2} \ln\dfrac{m^2}{\mu^2}{\Big )} \nn\\
& - \dfrac{b^2}{16 \pi^4}\sum_{n=1}^{\infty}\dfrac{(-1)^n}{n^2}{\Big (}  \exp(\dfrac{n \pi m^2}{b}) {\rm
Ei}(-\dfrac{n \pi m^2}{b}) \nn\\
& + \exp(-\dfrac{n \pi m^2}{b} ) {\Eibar}(\dfrac{n \pi m^2}{b} ){\Big)}\nn\\
& -i \dfrac{b^2}{16 \pi^3}\sum_{n=1}^{\infty}\dfrac{(-1)^n}{n^2}\exp(-\dfrac{n \pi m^2}{b}).
\eea
Now, it is clear that we can obtain one from the other by the dual transformation (11). So one need to know
only the effective action for the pure magnetic background to know the effective action for the pure electric
background, and vise versa. This demonstrates the power of the
duality.\par
In this paper we have established the
electric-magnetic duality in the quantum effective action of the standard QED and the scalar QED. From the
physical point of view the existence of the duality in the effective action is perhaps not so surprising. But the
fact that this duality is borne out from our calculation of one loop effective action is really remarkable. We expect
the duality to bean exact symmetry of the quantum effective action of gauge theories for an arbitrary
$n$-loop approximation.\par
One can show that exactly the same duality exists in non-Abelian gauge theories, in
particular in QCD \cite{cho2}. So it must be clear that the duality is a fundamental symmetry of the quantum
effective action of the gauge theories, both Abelian and non-Abelian. This means that we can use the duality as
a consistency check of the correctness of the quantum effective action of gauge theories.A more detailed
discussion on the subject will be published elsewhere
\cite{cho3}.\par
One of the authors (YMC) thanks Professor C.
N. Yang for the illuminating discussions.The work is supported in part by Korea Research Foundation
(KRF-2000-015-BP0072), and by the BK21 project of Ministry of Education.

\end{document}